\documentclass{moriond}

\usepackage{graphicx}
\usepackage{caption}
\usepackage{subcaption}
\usepackage{amsmath}

\def\be{\begin{equation}}
\def\ee{\end{equation}}
\def\bea{\begin{eqnarray}}
\def\eea{\end{eqnarray}}

\newcommand{\Photo}{\includegraphics[height=45mm]{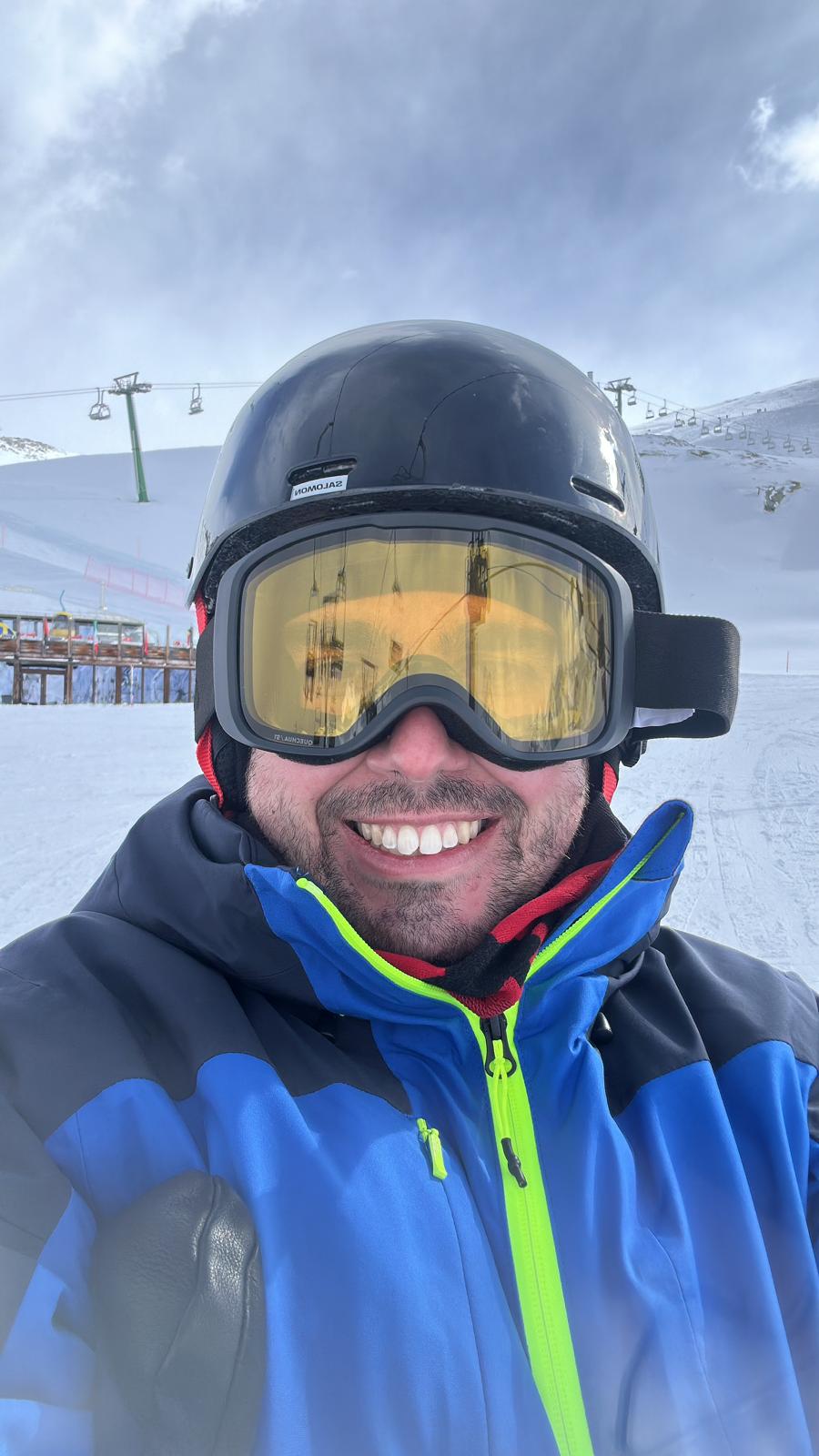}}

\begin{document}
\vspace*{4cm}
\title{QCD, electroweak physics, and searches for exotic signatures in the forward region at LHCb}

\author{Emilio X. Rodríguez Fernández}
\address{The Galician Institute for High-Energy Physics (IGFAE), Rúa de Xoaquín Díaz de Rábago, 15705 Santiago de Compostela, A Coruña (Spain) \\
\textit{On behalf of the LHCb collaboration}}

\maketitle 
\abstracts{The LHCb detector has demonstrated a proven competitiveness across a wide range of physics analyses thanks to its forward coverage. These proceedings describe: i) complementary measurements using heavy flavour jets, ii) Electroweak (EW) measurements with the top and W boson, and iii) searches for New Physics states such as axion-like particles (ALPs), heavy-neutral lepton (HNLs) and B-meson decays to multi-muon final states.}

\section{Introduction: the LHCb detector.}

The LHCb detector aims to contribute to the completion of the Standard Model (SM) from a wide range of perspectives while searching for potential deviations. As a forward single-arm spectrometer $(2 < \eta < 5)$, its exceptional tracking and vertexing performance, combined with powerful particle identification (PID) and operations at low pile-up $(\langle \mu \rangle \approx 1.1)$, enabled the precise reconstruction of complex topologies with a two-level trigger system, composed by a Hardware (L0) and Software (HLT1 and HLT2) levels during Run 1 and 2 of the LHC. This strategy focused on precision rather than abundance, supporting the LHCb program not only in high-precision measurements in the QCD and electroweak sectors but also for capturing the low-mass and displaced signatures characteristic of Beyond the Standard Model (BSM) phenomena.

\begin{figure}
    \centering
\begin{subfigure}[b]{0.4\linewidth}
        \includegraphics[width=\linewidth]{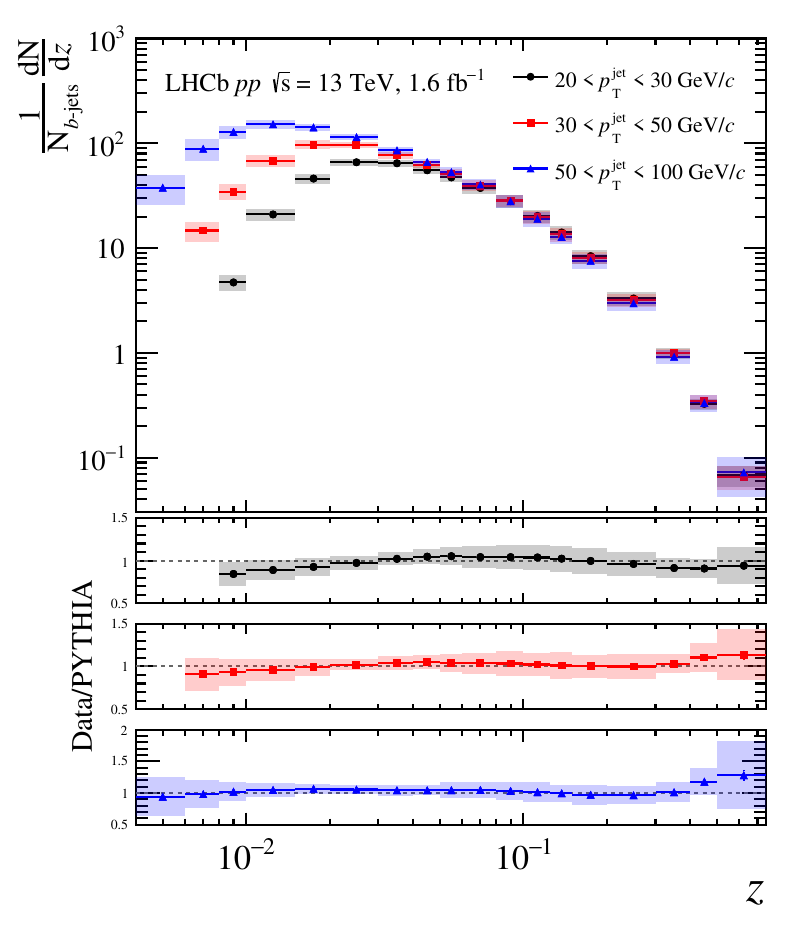}
        \subcaption{Number of heavy flavour jets as a function $z$ for different pT ranges~\cite{LHCb:2025zmu}.}
    \end{subfigure}
    \begin{subfigure}[b]{0.4\linewidth}
        \includegraphics[width=\linewidth]{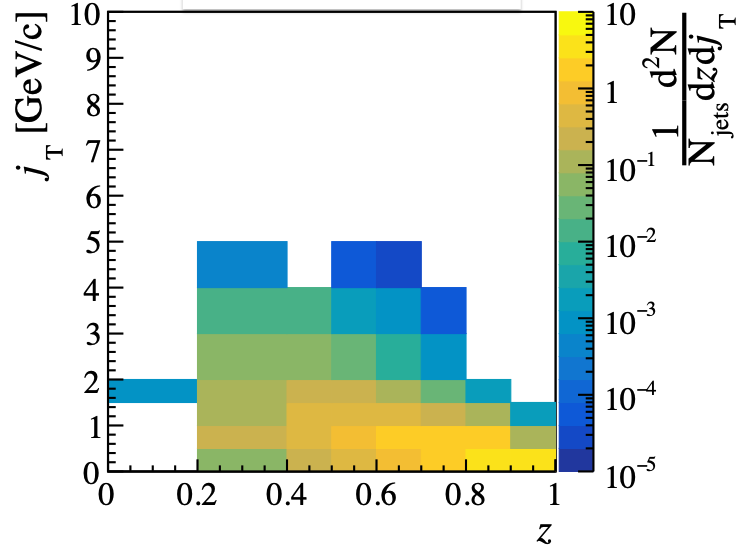}
        \subcaption{$j_T$ vs $z$ plot for b-jet fragmentation in $B^{\pm} \to J/\psi (\mu^+ \mu^-) K^{\pm}$ decays~\cite{LHCb:2026mbl}.}
    \end{subfigure}
    \caption{}
    \label{fig:jets}
\end{figure}

\section{Jet measurements}

The LHCb detector's pseudorapidity coverage enables detailed studies of small- and high-Bjorken-$x$ parton distributions (PDFs), complementing central-rapidity measurements. 

LHCb has studied the longitudinal momentum fraction $z$, transverse momentum $j_T$, and radial distribution $r$ of charged particles in heavy-flavour jets. Jets are reconstructed with the anti-$k_T$ algorithm, with additional selections for combinatorial background and high-energy leptons. Heavy-flavour jets are tagged using the Secondary Vertex (SV) algorithm~\cite{LHCb2015btag} and a BDT classifier. Corrections for trigger, tracking, PID mismodelling and a Bayesian Unfolding technique (correct measured detector-level distributions for finite resolution, inefficiencies, and migration effects to recover the true particle-level distributions) are applied. Results for charged-hadron $z$ in different $p_T$ bins are shown in Figure~\ref{fig:jets}a. Similarly, $b$-quark PDFs with $B^+ \to J/\psi(\mu^+ \mu^-) K^{\pm}$ were studied, restricting to collision events with a single $pp$ interaction. As an example, the 2D PDF in the plane $(j_T, z)$ is displayed in Figure~\ref{fig:jets}b. 

Alternative methods using a Gradient Boosting Regressor for energy calibration and a Deep Neural Network for flavour tagging have been tested with inclusive $H \to b\overline{b}, c\overline{c}$ events~\cite{LHCb:2026ezb}. The mass spectrum is dominated by multijet QCD and $Z \to b\overline{b}, c\overline{c}$, yielding limits $\sigma_{H \to b\overline{b}} = 11.1 \, \sigma^{\rm{SM}}_{H \to b\overline{b}}$ and $\sigma_{H \to c\overline{c}} = 1834 \, \sigma^{\rm{SM}}_{H \to c\overline{c}}$, constraining the charm Yukawa coupling to $y_c = 43 \, y^{\rm{SM}}_c$~\cite{LHCb:2026ezb}.

\section{Electroweak measurements}

\begin{figure}
    \centering
    \begin{subfigure}[b]{0.4\linewidth}
        \includegraphics[width=\linewidth]{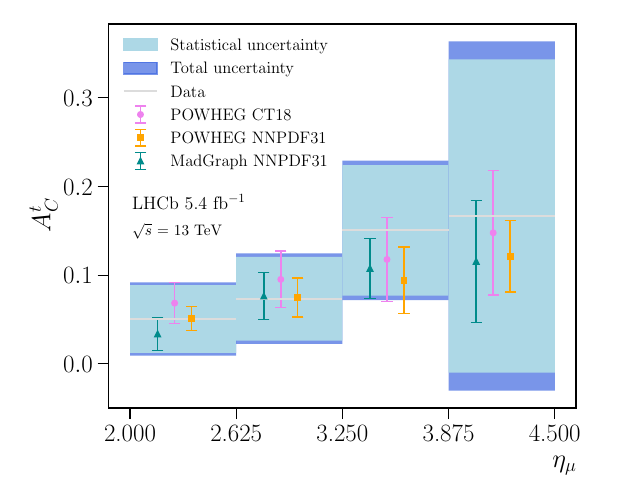}
        \subcaption{$t \overline{t}$ production charge asymmetry~\cite{LHCb:2025kfp}.}
    \end{subfigure}
    \begin{subfigure}[b]{0.4\linewidth}
        \includegraphics[width=\linewidth]{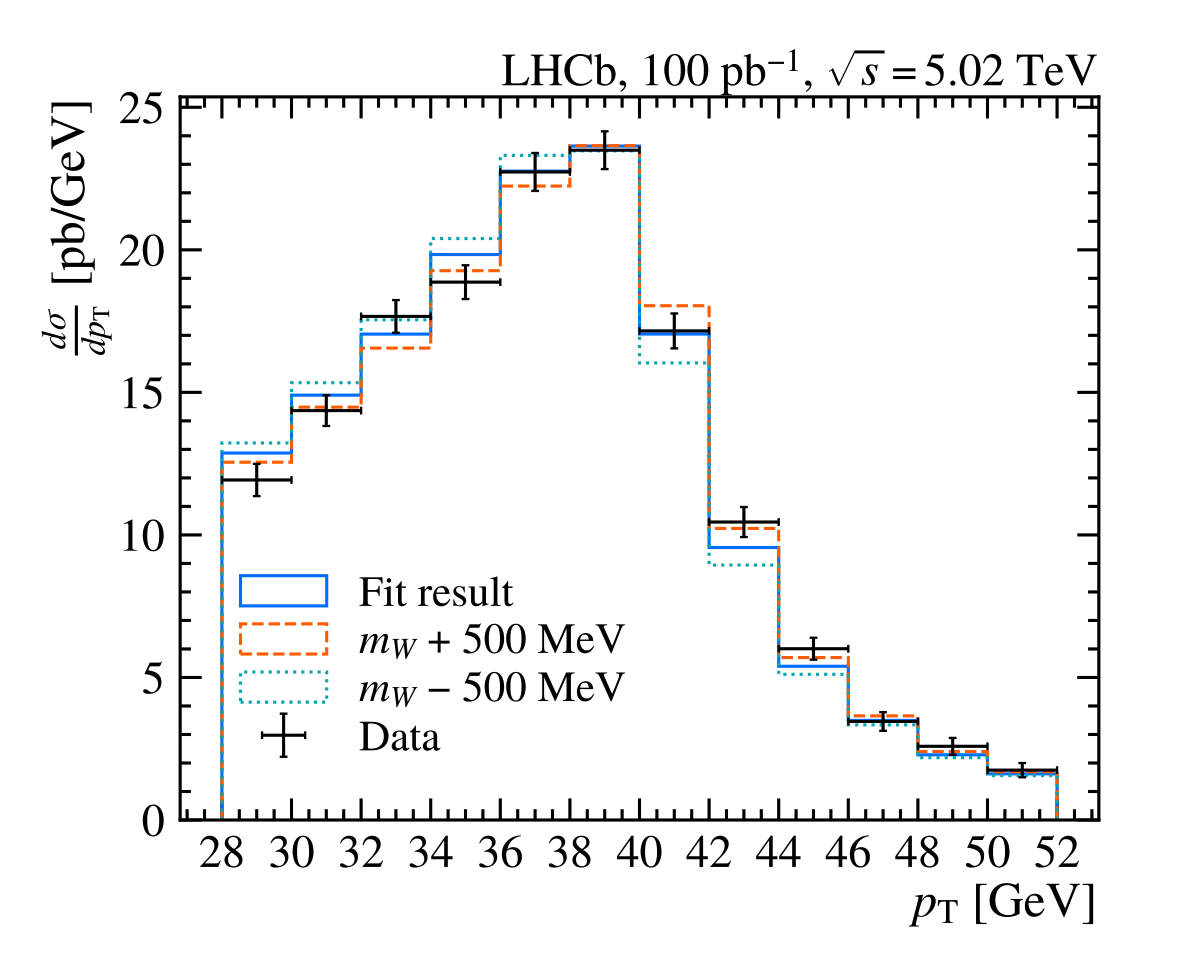}
        \subcaption{Unfolded $d\sigma/dp_T$ for $W^+$ boson~\cite{LHCb:2025msn}.}
    \end{subfigure}
    \caption{}
    \label{fig:EW}
\end{figure}

LHCb's forward geometry also helps in complementing other precision studies of EW parameters, which aim to test the SM to its limits. Figure~\ref{fig:EW}a shows the $t\overline{t}$ production charge asymmetry and $W$ boson mass measurements. LHCb measured the charge asymmetry in $t\overline{t}$ production from $t \to W^+(\mu^+\nu_\mu)b$ decays using a DNN for jet-flavour identification. Backgrounds are reduced by: i) $m(\mu^+\mu^-) < 40\,\rm{GeV}$ for $Z/\gamma^*(\mu^+\mu^-)b$, and ii) $p_T^\mu > 20\,\rm{GeV}$ with high muon isolation for multijet QCD. The integrated results are $\sigma_{t} = 0.95 \pm 0.04 \pm 0.08 \pm 0.02 \, \mathrm{pb}$ and $\sigma_{\overline{t}} = 0.81 \pm 0.03 \pm 0.07 \pm 0.02 \, \mathrm{pb}$;
where the last error is luminosity uncertainty. LHCb also measured the $W \to \mu\nu_\mu$ cross-section and the $W$ boson mass~\cite{LHCb:2025msn} in bins of muon $p_T$ and isolation, using templates for signal and backgrounds (high-$p_T$ misidentified hadrons, EW with $\tau$'s from $W/Z$). Backgrounds like $Z \to \mu^+\mu^-$ or muonic decays of long-lived hadrons are subtracted by discarding a second muon with $p_T > 25\,\rm{GeV}$ and requiring isolation $I^{\mu} < 8\,\rm{GeV}$. The measurement uses $100\,\rm{pb^{-1}}$ at $\sqrt{s}=5.02\,\rm{TeV}$, with the unfolded $d\sigma/dp_T$ distribution for the $W^+$ boson being shown in Figure~\ref{fig:EW}b.

\section{Exotic signatures in the forward region}

The unique LHCb properties also enable BSM searches via novel strategies, i.e., di-photon signatures and displaced vertexing, for which information from all subdetectors (Long tracks) or from those located downstream of the VELO (Downstream tracks) is used.

\begin{figure}[ht]
    \centering
    \begin{subfigure}[m]{0.4\linewidth}
        \includegraphics[width=\linewidth]{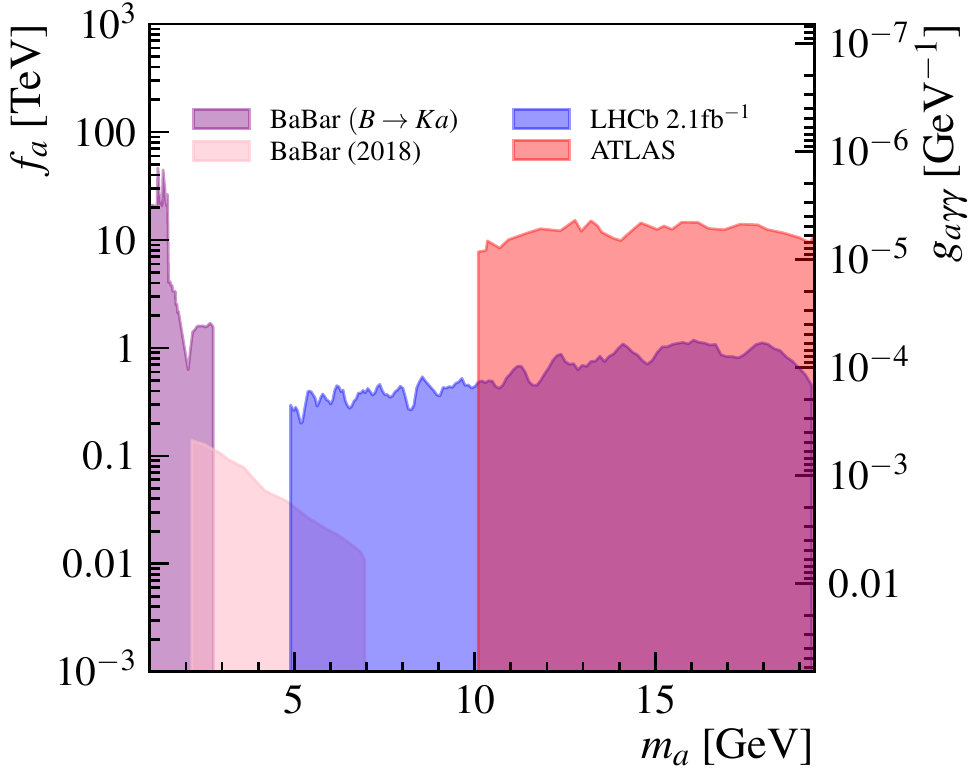}
        \subcaption{State of the art concerning $\rm{ALP}$ searches~\cite{LHCb:2025gbn}.}
    \end{subfigure}
    \begin{subfigure}[m]{0.4\linewidth}
        \includegraphics[width=\linewidth]{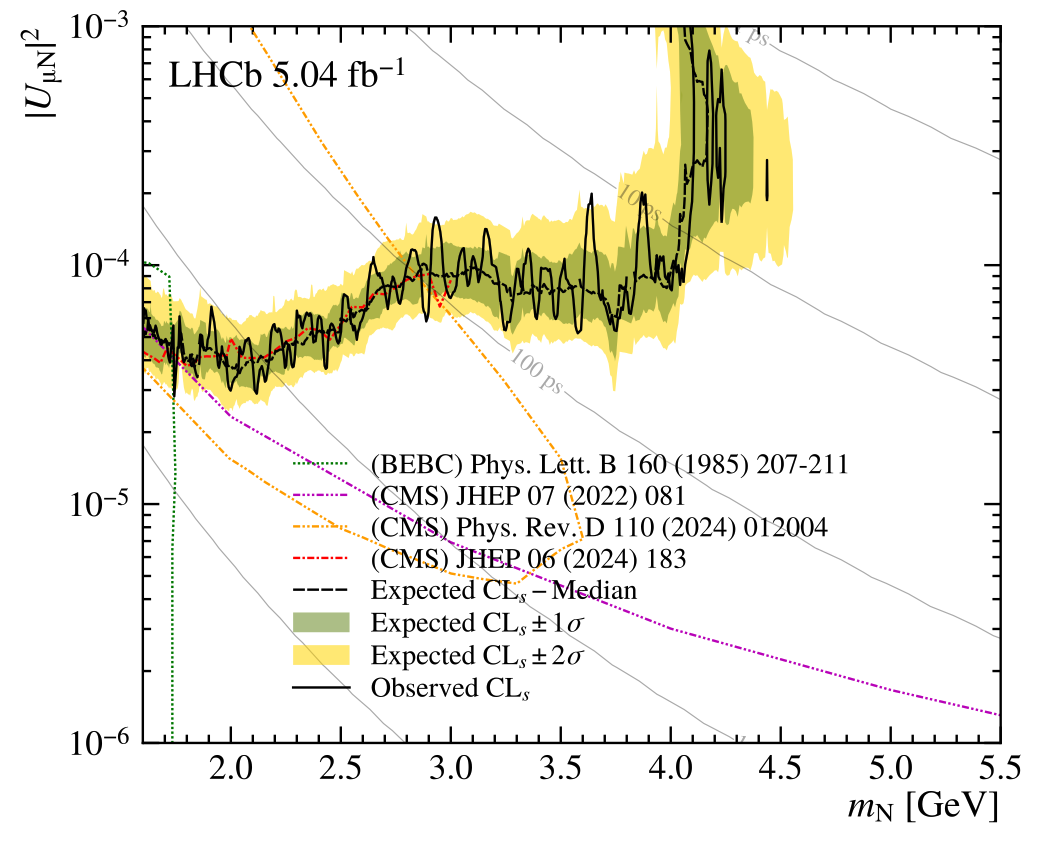}
        \subcaption{HNL exclusion limits~\cite{LHCb:2025ymr}.}
    \end{subfigure}
    \begin{subfigure}[m]{0.4\linewidth}
        \includegraphics[width=\linewidth]{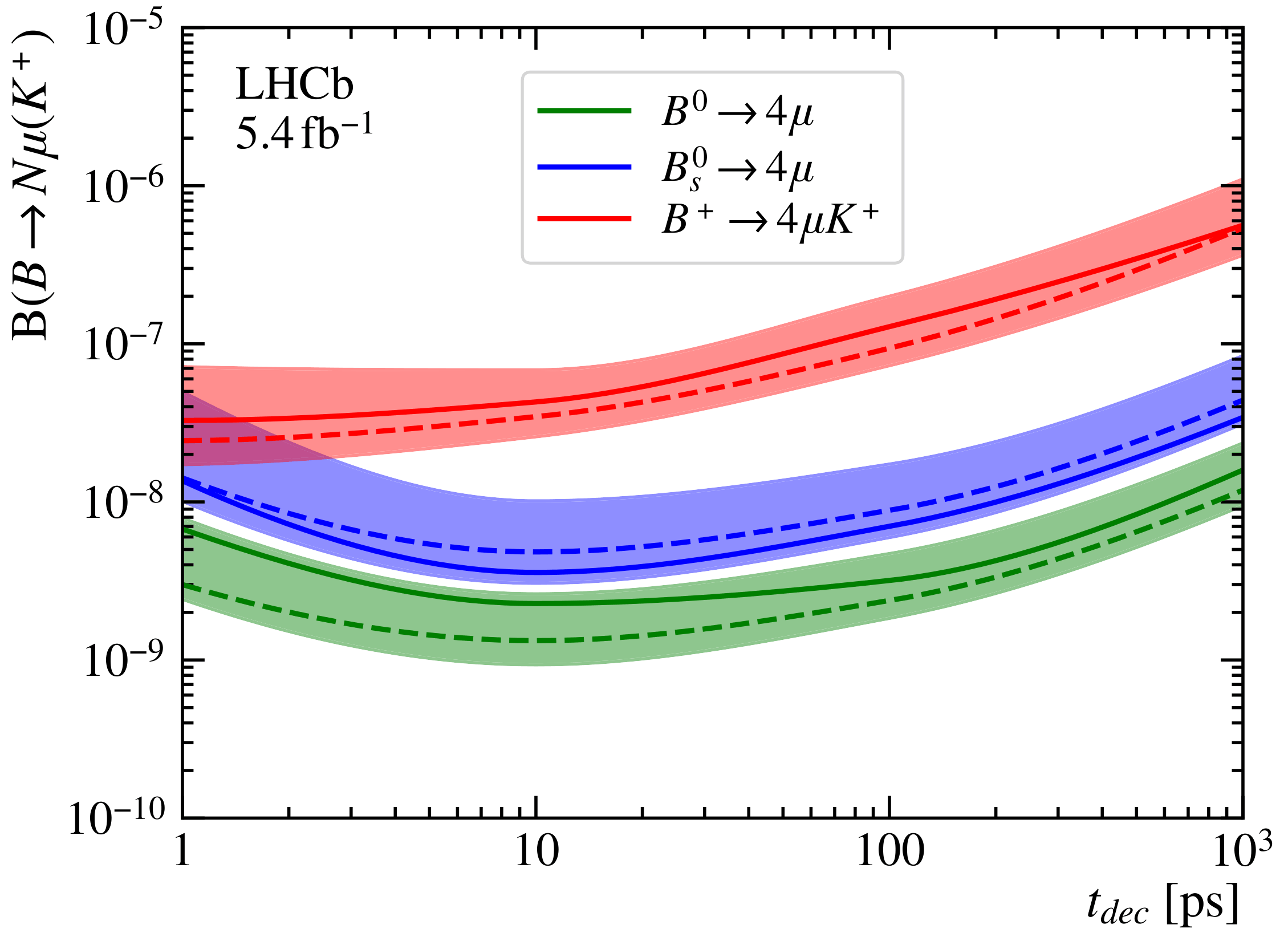}
        \subcaption{Lifetime-dependent limits for $B_{u,d,s}\to 4\mu (K^+)$ decays \cite{LHCb-PAPER-2026-014}.}
    \end{subfigure}
    \caption{}
    \label{fig:NP}
\end{figure}

Axion-like particles (ALPs) coupling to gluons are produced via gluon fusion and reconstructed as $a\to\gamma\gamma$~\cite{LHCb:2025gbn}. Using $2.1$~fb$^{-1}$ of Run~2 data, LHCb searches in the range $m_{\gamma\gamma}\in[4.9,19.4]$~GeV. Photon pairs are reconstructed in ECAL with high-$E_T$ clusters consistent with the primary vertex, selected via a multivariate algorithm (isolation variables) with ECAL saturation veto and photon identification. The main background $B^0 \to \pi^0\pi^0$ is suppressed with PID cuts. Stringent limits are set for prompt ALPs in $m_a\in[4.9,10]$~GeV (Figure~\ref{fig:NP}a). 

Heavy neutral leptons (HNLs) couple via the lepton-Yukawa portal or higher-dimensional operators, possibly with a Majorana mass term~\cite{LHCb:2025ymr}. Using full Run~2, LHCb searches for HNLs from $B$-meson decays, scanning $m_N \in [1.6, 5.5]$~GeV. Events are categorized by reconstruction (Long/Downstream) and topology (same-sign dimuons for Majorana, opposite-sign for Dirac). A neural network suppresses combinatorial background. Figure~\ref{fig:NP}b shows $95\%$ CL limits on $|U_{\mu N}|^2$ in $(m_N, t_N)$, with LHCb providing the most stringent constraints at large masses and short lifetimes. 

In the $B$-meson sector, LHCb also analyzed multimuon decays~\cite{LHCb-PAPER-2026-014} arising through pseudo Nambu-Goldstone bosons ($a_1$, $a_2$), predicted in Supersymmetry and Composite Higgs models. The analysis searches for $B_s \to (4,6)\mu$ and $B^+ \to (4,6)\mu K^+$ with prompt or displaced mediators. Corrections for simulation mismodeling are performed, and selections include: i) mass vetoes (prompt) or displacement requirements (displaced), ii) multivariate selections for $4\mu$ modes, iii) mass resolution criteria for $6\mu$ modes. Results are normalized to $B_s^0 \to J/\psi(\mu^+\mu^-)\phi(\mu^+\mu^-)$ to cancel systematic uncertainties. Limits for $B_{u,d}$ decays with $m_{a_1}=0.25\,\rm{GeV}$, $m_{a_2}=0.40\,\rm{GeV}$ are shown in Figure~\ref{fig:NP}c.

\section{LHCb Upgrade 1}

The LHCb Upgrade I introduces significant improvements: increased luminosity conditions with $\langle \mu \rangle = 5.2$, requiring upgrades to tracking subdetectors (VELO and replacement of inner/outer trackers with the SciFi detector), enhanced PID capabilities, and removal of the L0 hardware trigger in favor of a full software trigger enabling higher-rate data acquisition. These bring clear improvements on several fronts: inclusive analysis of $H \to b\bar{b}, c\bar{c}$, with a projection of $\sigma_{H \to b\bar{b}} = 0.38 , \sigma^{\rm{SM}}$, $\sigma_{H \to c\bar{c}} = 45, \sigma^{\rm{SM}}$, and $y_c = 6.7 , y^{\rm{SM}}_c$ \cite{LHCb:2026ezb}; electroweak measurements, with a clear reduction on the statistical error on $t\bar{t}$ charge asymmetry and $W$ mass; and BSM searches, with dielectron modes benefiting from looser thresholds and PID (also increasing dimuon statistics) plus alternative standalone reconstructions with tracking stations downstream of the magnet and improved vertexing for displaced physics analyses.

\section*{References}
\bibliography{moriond.bib}

\end{document}